\def\year{2017}\relax
\documentclass[letterpaper]{article} 
\usepackage{aaai17}  
\usepackage{times}  
\usepackage{helvet}  
\usepackage{courier}  
\usepackage{url}  
\usepackage{graphicx}  

\usepackage{url}
\usepackage{amsmath}
\usepackage{subfigure}
\usepackage{multirow}
\usepackage{ctable}
\usepackage{booktabs}

\newcommand{\squishlist}{
\begin{list}
	{$\bullet$} { \setlength{
	\itemsep}{0pt} \setlength{\parsep}{3pt} \setlength{\topsep}{3pt} \setlength{
	\partopsep}{0pt} \setlength{\leftmargin}{1.5em} \setlength{\labelwidth}{1em} \setlength{\labelsep}{0.5em} } }
	
	\newcommand{\squishlisttwo}{
	\begin{list}
		{$\bullet$} { \setlength{
		\itemsep}{0pt} \setlength{\parsep}{0pt} \setlength{\topsep}{0pt} \setlength{
		\partopsep}{0pt} \setlength{\leftmargin}{2em} \setlength{\labelwidth}{1.5em} \setlength{\labelsep}{0.5em} } }
		
		\newcommand{\squishend}{
	\end{list}
	}

\begin{document}
%
\title{Characteristics of On-time and Late Reward Delivery Projects}
\author{
 Thanh Tran \\ Department of Computer Science \\ Utah State University, Logan, UT 84322 \\ thanh.tran@aggiemail.usu.edu \And
 Kyumin Lee \\ Department of Computer Science \\ Utah State University, Logan, UT 84322 \\ kyumin.lee@usu.edu}
\maketitle
\begin{abstract}
The crowdfunding market size has increased exponentially, reaching tens of billions of dollars and showing the popularity of crowdfunding. However, according to Kickstarter, 35\% backers did not receive rewards on time. To maintain the trust between creators and backers, and sustain the crowdfunding business growth, it is crucial to understand how on-time and late reward delivery projects are different. In this paper, we analyze characteristics of on-time and late reward delivery projects, especially, focusing on project descriptions, creator profiles, and activeness and linguistic patterns of creators and backers. Our analysis reveals that the less complicated a project is and more actively a creator responds to backers, the higher an on-time reward delivery probability is. It shows there are significant differences between on-time and late reward delivery projects.



\end{abstract}

\section{Introduction}
\begin{figure*}
	\centerline{
		\includegraphics[width=0.8\linewidth]{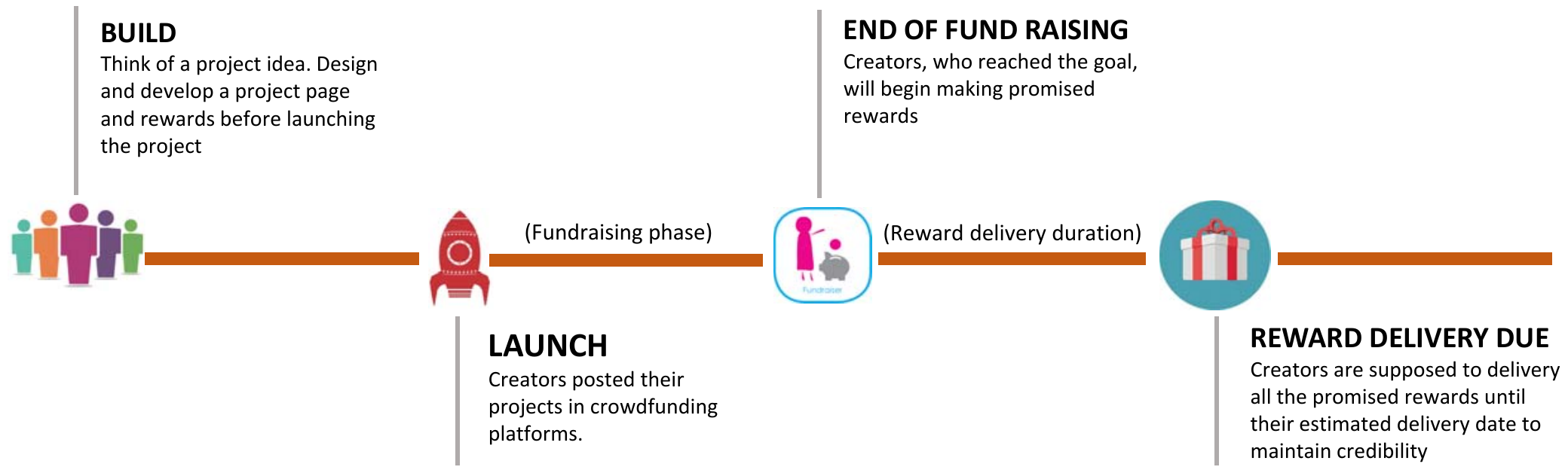}
	}
	\caption{Project Timeline in reward-based crowdfunding platforms.}
	\label{fig:campaignTimeline}
\end{figure*}

In recent years, reward-based crowdfunding platforms like Kickstarter and Indiegogo have become popular with increasing number of projects and backers. As shown in Figure \ref{fig:campaignTimeline}, a project has two phases: \emph{the fundraising phase} and \emph{the reward delivery phase}. In \emph{the fundraising phase}, creators seek funding for their projects, and backers fund projects that they are interested in. In \emph{the reward delivery phase}, creators of successfully funded projects make and ship their rewards.




In the literature, researchers mostly focused on \emph{the fundraising phase} by understanding the dynamics of crowdfunding platforms as well as the interaction and the relationship between creators and backers \cite{kuppuswamy2015crowdfunding,tran2016succeed}, predicting successful projects \cite{chung2015long,li2016project}, and recommending creators to backers or vice versa \cite{rakesh2015project}. 

However, little attention has been paid to \emph{the reward delivery phase}, even though 35\% backers did not receive rewards on time\footnote{https://www.kickstarter.com/fulfillment}. If receiving rewards on time, the backers will be more likely to invest the creator's new projects. Therefore, delivering the rewards on time is crucial as it helps creators to maintain trust with backers and to retain backers' upcoming investments.

To achieve trustworthy crowdfunding platforms, the first step is to understand how on-time and late reward delivery projects are different. Can we find distinguishing properties between on-time and late reward delivery projects? In this paper, we focus on this question and analyze characteristics of these projects. If we find distinguishing properties, creators can improve new projects' descriptions, and communication skills with backers toward planning their project well and delivering promised rewards on time.

In particular, we extracted and analyzed various characteristics of projects and creators which distinguish between on-time and late reward delivery projects. Through our comprehensive analysis of 2,198 sampled Kickstarter projects, we found that:


\begin{itemize}
    \item There were 12 project and creator related features, which distinguished between on-time and late reward delivery projects. Less complicated projects (e.g., a lower amount of goal and less number of rewards) had a higher chance to deliver rewards on time. 
                
    \item During \emph{the reward delivery phase}, creators in on-time reward delivery projects posted updates and comments more quickly than creators in late reward delivery projects. Backers in on-time reward delivery projects asked less number of questions through comments.
        
    \item Creators in on-time and late reward delivery projects had different linguistic usage in their updates. Similarly, we found that backers in on-time and late reward delivery projects had different linguistic usage in their comments.
\end{itemize}

\section{Dataset}
\label{sec:dataset}
\begin{figure} [t]
	\centering
	\includegraphics[width=0.2\textwidth]{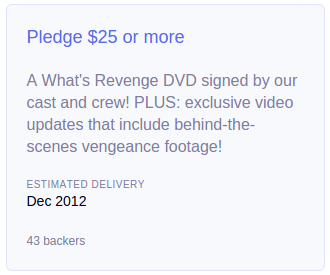}
	\caption{An example of a Kickstarter reward.}
	\label{fig:RewardExample}
\end{figure}

Figure \ref{fig:RewardExample} shows a reward example on Kickstarter. A reward consists of a description, price, the number of pledged backers, and an estimated delivery date. According to Kickstarter policy, a creator sets up her rewards before launching the project. Once a reward is pledged, it is not allowed her to edit the rewards. Based on this, we first define what on-time and late reward delivery projects mean as follows:
\begin{itemize}
	\item A project is called an on-time reward delivery project if all the rewards associated with the project were shipped by the longest estimated delivery date (LEDD) among estimated delivery dates of the rewards.
	\item A project is called a late reward delivery project if the creator did not deliver at least one of rewards by the longest estimated delivery date (LEDD).
\end{itemize}

First of all, we selected 29,499 successful projects launched between 2009 and September 2014, and their goals were at least \$100. 
Next, we sampled 10\% of 29,499 successful projects with keeping the same project distribution over project categories, year and goal. Finally, we sampled 2,949 projects, and collected their pages, corresponding creator profiles, and associated updates and comments. To get the ground-truth regarding whether a project is on-time or late reward delivery project, three labelers independently read all the project updates and comments and labeled the 2,949 projects based on the following guidelines:
\begin{itemize}
	\item A project was labeled as an on-time reward delivery project if: (i) the labeler could identify that all the rewards were shipped before LEDD, and (ii) there was no complaint regarding not receiving the rewards.
	\item A project was labeled as a late reward delivery project if the creator posted at least one update or comment saying that the reward shipping process would be delayed and a new delivery date beyond LEDD was provided.
	\item A project would be excluded if there was no information to verify whether all the rewards were sent on time or not.
\end{itemize}

Finally, 1,003 projects were labeled as on-time reward delivery projects, and 1,195 projects were labeled as late reward delivery projects. 751 projects were excluded because of the third guideline. 


\section{Characteristics}

In this section, we analyze the 2,198 projects, and associated creators and backers' behavior to understand characteristics of on-time and late reward delivery projects.

\begin{figure}[t]
	\centering
	\includegraphics[width=0.3\textwidth]{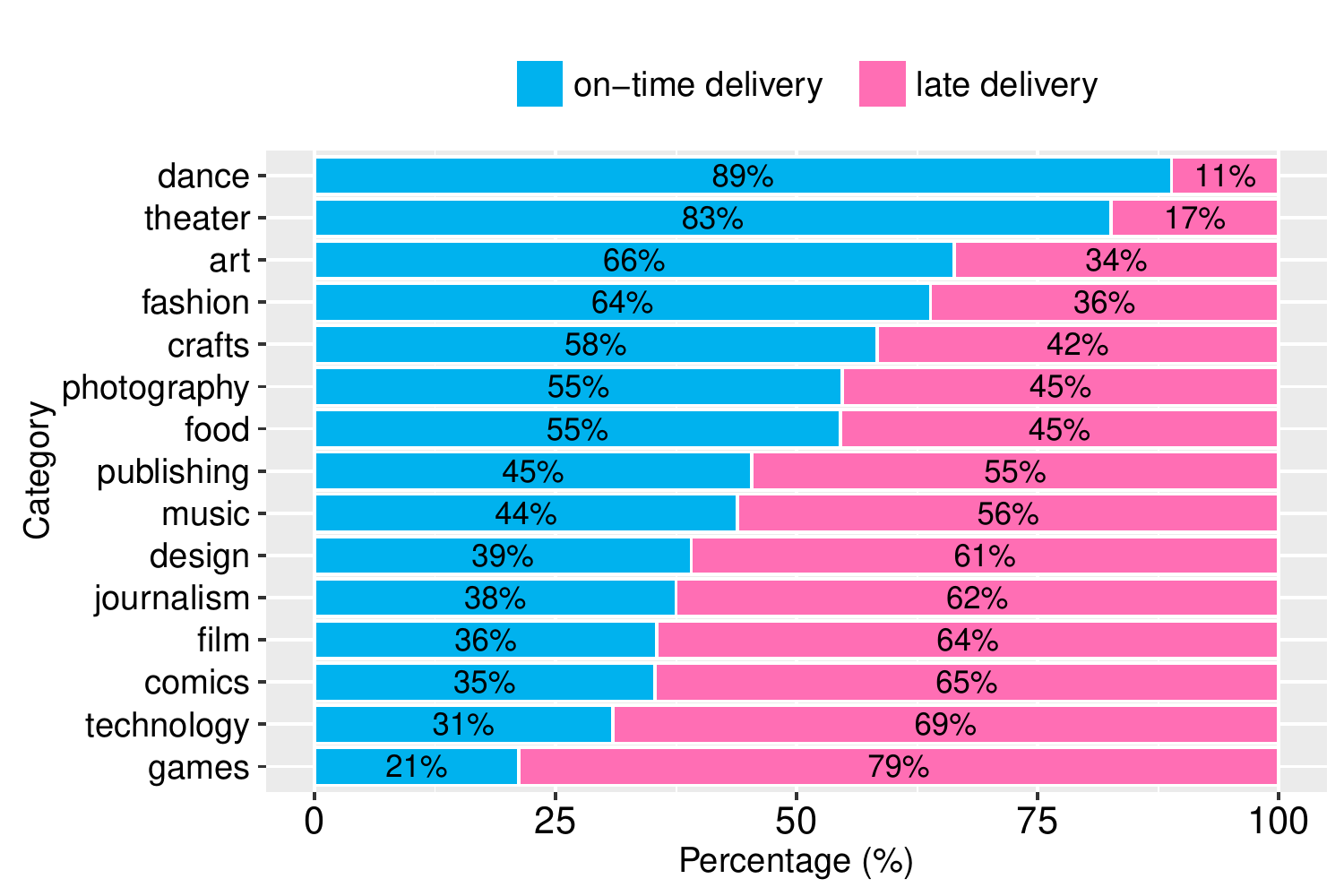}
	\caption{Category distributions of on-time and late reward delivery projects.}
	\label{fig:OntimePassedProjectsDistribution}
\end{figure}
\begin{figure}[t]
	\centering
	\includegraphics[width=0.3\textwidth]{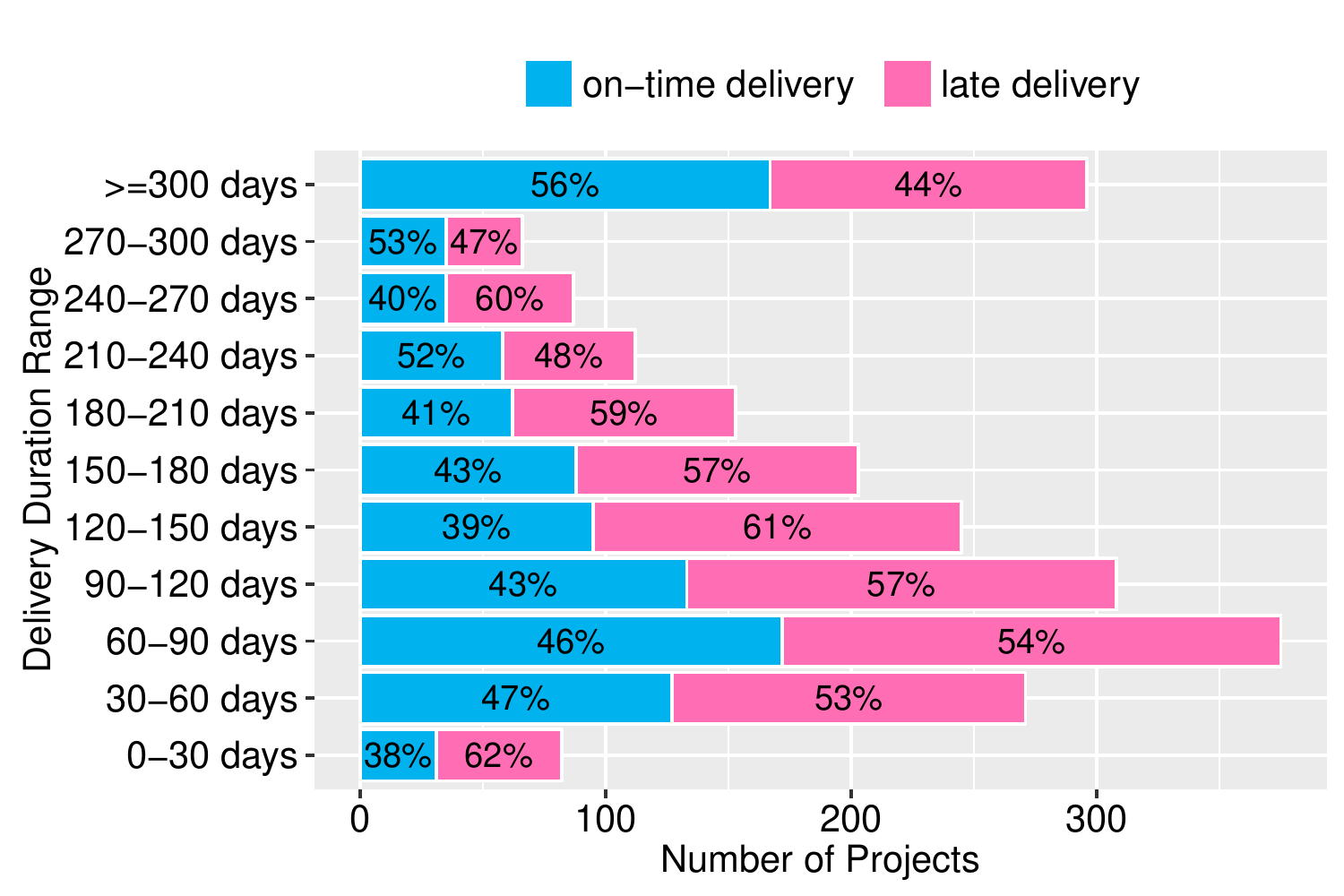}
	\caption{Distribution of on-time and late delivery projects in various delivery duration ranges.}
	\label{fig:OntimePassedProjectsDistributionOverDeliveryDuration}
\end{figure}
\begin{table*} [t]
	\caption{Characteristics of on-time and late reward delivery projects . $std$ indicates standard deviation; *,**,***, ns indicates p-value $< 0.05$, $< 0.01$, $< 0.001$ and \emph{not significant}, respectively.
		\label{table:project-charac}}
	\centering
	{
		\resizebox{0.8\textwidth}{!}{
			\centering
			\small
			\begin{tabular}{lrrrrrrr}
				& \textbf{on-time delivery} 	& $std$ 	&\textbf{late delivery}&	$std$ 	& \textbf{Total} & $std$ & \textbf{p-value} \\
				\toprule
				Avg. Fundraising Duration (days) 	& 32.13 	& 10.13 	& 33.71 		& 10.48 	& 32.99 	&  10.35	& ***	\\
				Avg. Goal (USD) 				& 7,670.7 	& 29,888.04	& 13,584.48 	& 24,652.49	& 10,885.88 & 27,326.27	& ***	\\
				Avg. pledged money (USD)		& 14,419.15 & 134,394.58& 30,126.76 	& 128,016.97& 22,959 	& 131,199.24& **	\\
				Avg. \# of images 					& 3.94 		& 7.39 		& 5.72 			& 9.45  	& 4.9 		& 8.62		& ***	\\
				Avg. \# of videos 					& 1.23 		& 0.95		& 1.38 			& 1.16  	& 1.31 		& 1.07		& ***	\\
				
				Avg. \# of FAQs 					& 0.85 		& 1.90 		& 1.46 			& 2.93  	& 1.18 		& 2.53		& ***	\\
				Avg. \# of rewards 					& 10.09		& 5.18 		& 11.52 		& 6.61  	& 10.87		& 6.04		& ***	\\
				Avg. \# of backers 					& 234.58 	& 2408.07	& 395.32 		& 2662.73  	& 321.97 	& 2550.94	& ns	\\

				Avg. \# of sentences in a project desc	& 12.50		& 2.58		& 12.31 		& 2.09 		& 12.40 	& 2.33		& *		\\
				Avg. \# of sentences in a reward desc 	& 21.58 	& 22.64		& 25.68 		& 25.81 	& 23.81 	& 24.5		& ***	\\
				Avg. smog score of a project desc 	& 30.12 	& 24.51		& 38.59 		& 34.22 	& 34.72 	& 30.47		& ***	\\
				Avg. smog score of a reward desc 	& 12.77 	& 5.98 		& 12.65 		& 5.25  	& 12.70	 	& 5.59		& ns	\\
				\bottomrule
			\end{tabular}
		}
	}
\end{table*}

\begin{table*}[t]
	\caption{Characteristics of creators in on-time and late reward delivery projects . $std$ indicates standard deviation; *,**,***, ns indicate p-value $< 0.05$, $< 0.01$, $< 0.001$ and \emph{not significant}, respectively.
		\label{table:creator-charac}}
	\centering
	{
		\resizebox{0.8\textwidth}{!}{
			\centering
			\small
			\begin{tabular}{lrrrrrrr}
				& \textbf{on-time delivery} 	& $std$ 	&\textbf{late delivery}&	$std$ 	& \textbf{Total} & $std$ & \textbf{p-value} \\
				\toprule
				Avg. \# of backed projects 		& 7.06 	& 17.11 	& 9.69 		& 19.74 	& 8.49 	&  18.63	& ***	\\
				Avg. \# of created projects		& 1.65 	& 3.87		& 1.75		& 3.26		& 1.70	&  3.55		& ns 	\\
				Avg. \# of sentences in a bio desc   & 6.36	& 7.34		& 6.09		& 6.04		& 6.21	&  6.66		& ns	\\
				Avg. smog score of a bio desc  & 13.59	& 3.29		& 13.24		& 3.59		& 13.4	&  3.46		& * 	\\
				Avg. \# of Facebook friends  	& 690.8	& 973.01	& 637.7		& 904.18	& 661.93&  936.59	& ns 	\\
				\bottomrule
			\end{tabular}
		}
	}
\end{table*}

\smallskip
\noindent\textbf{Project Characteristics:} When creating a project, the creator has to select its category from 15 categories predefined by Kickstarter. We are interested in finding whether a proportion of on-time reward delivery projects was equal or not across the 15 categories. Can we observe creators in certain categories sent more rewards on time? To answer the question, we analyzed category distributions and plotted them in Figure~\ref{fig:OntimePassedProjectsDistribution}. Interestingly, rewards in most dance and theater related projects were delivered on time, whereas rewards in most games, technology, comics, and film related projects were not delivered on time. We analyzed these rewards, and found that project rewards in dance and theater were live performances, showcases or teaching dancing classes which were served to backers at once. In contrast, project rewards in other categories such as games, technology, comics, and film were real products (e.g., a game, book, movie), requiring more time to produce and ship the products to backers. The analysis and observation confirm that creators in certain categories delivered more rewards on time.

We next analyzed project related features to see if they help to distinguish on-time and late reward delivery projects. Table \ref{table:project-charac} presents statistical information of 12 project features: fundraising duration, goal, amount of pledged money, \# of images, \# of videos, \# of FAQs, \# of rewards, \# of backers, \# of sentences in a project description, \# of sentences in a reward description, and a smog score \cite{mclaughlin2008smog} of each of project and reward descriptions. We performed the one-tailed two samples t-test to determine which feature had a significant mean difference in one direction. We found that on-time reward delivery projects had some distinguishing characteristics compared to late reward delivery projects: (i) shorter fundraising duration; (ii) lower goal and pledged money; (iii) less number of images, videos, FAQs, and rewards; (iv) larger number of sentences in the project description, but smaller number of sentences in the reward description; and (v) lower smog score of a project description, indicating more readable.


We are also interested in exploring ``Did a longer delivery duration range increase a proportion of on-time delivery projects?''. Figure~\ref{fig:OntimePassedProjectsDistributionOverDeliveryDuration} shows a proportion of on-time and late delivery projects in each delivery duration range with the number of projects. It shows that the proportion of on-time delivery projects was quite equally distributed across delivery duration ranges. Even 44\% projects, whose duration was greater than 10 months, were late delivery projects. It indicates that longer delivery duration does not always increase a proportion of on-time delivery projects.

\smallskip
\noindent\textbf{Creator Characteristics:} We extracted 5 features related to creators: \# of backed projects, \# of created projects, \# of sentences in a bio description, smog score of a bio description, and \# of Facebook friends. Then, we performed one-tailed two samples t-test via the same process as we did for the project related features. Table \ref{table:creator-charac} shows our statistical results for those features. There were two features differentiated on-time and late reward delivery projects: \# of backed projects and smog score of a bio description. We observed that creators in on-time reward delivery projects had a smaller number of backed projects and a higher smog score.

\begin{figure*} [t]
\centering
	\subfigure[The average update time interval (***).] 
	{
		\label{fig:HowFast-Creator-Update}
		\includegraphics[width=0.28\textwidth, height=1.1in]{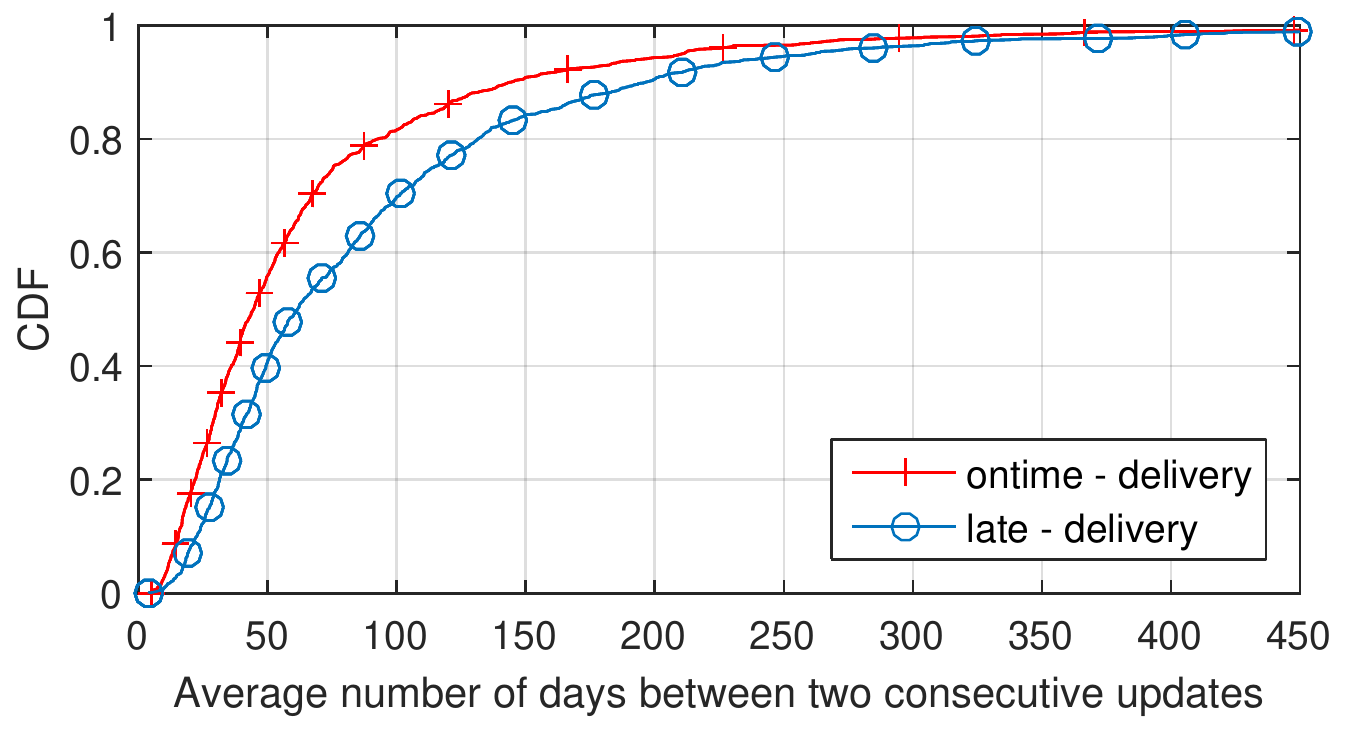}
	}
	\hspace{0.05cm}
	\subfigure[The average time between a backer's question and reply from a creator. (*). ] 
	{
		\label{fig:HowFast-Creator-Comment}
		\includegraphics[width=0.28\textwidth, height=1.1in]{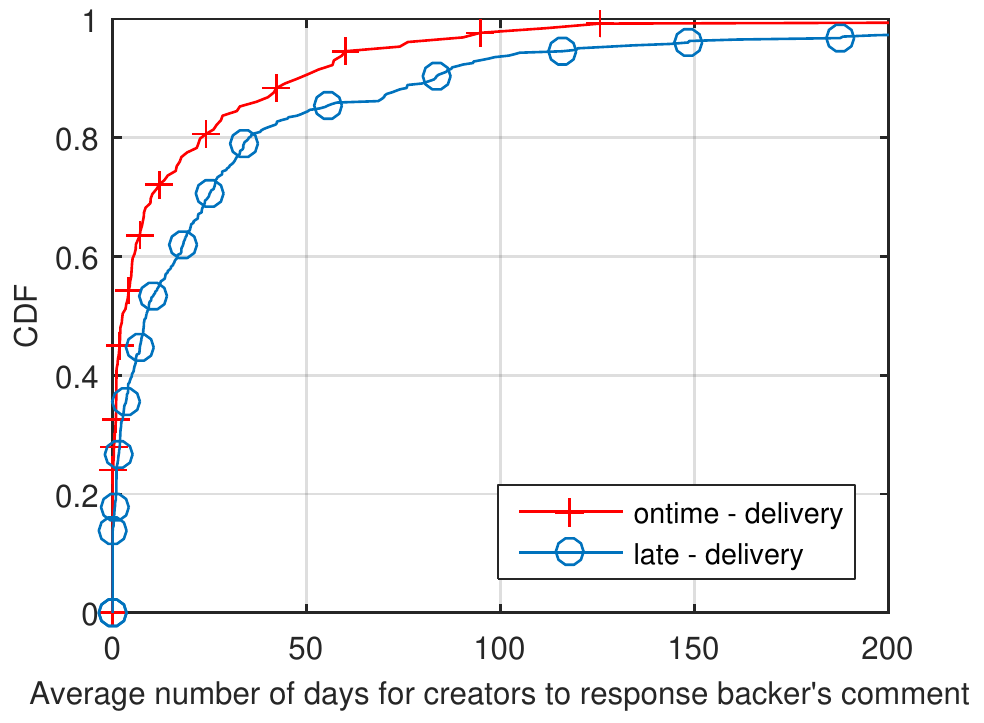}
	}
	\hspace{0.05cm}
	\subfigure[The number of questions posted by backers (***). ] 
	{
		\label{fig:CDF-BackerAsk4RewardProgress}
		\includegraphics[width=0.28\textwidth, height=1.1in]{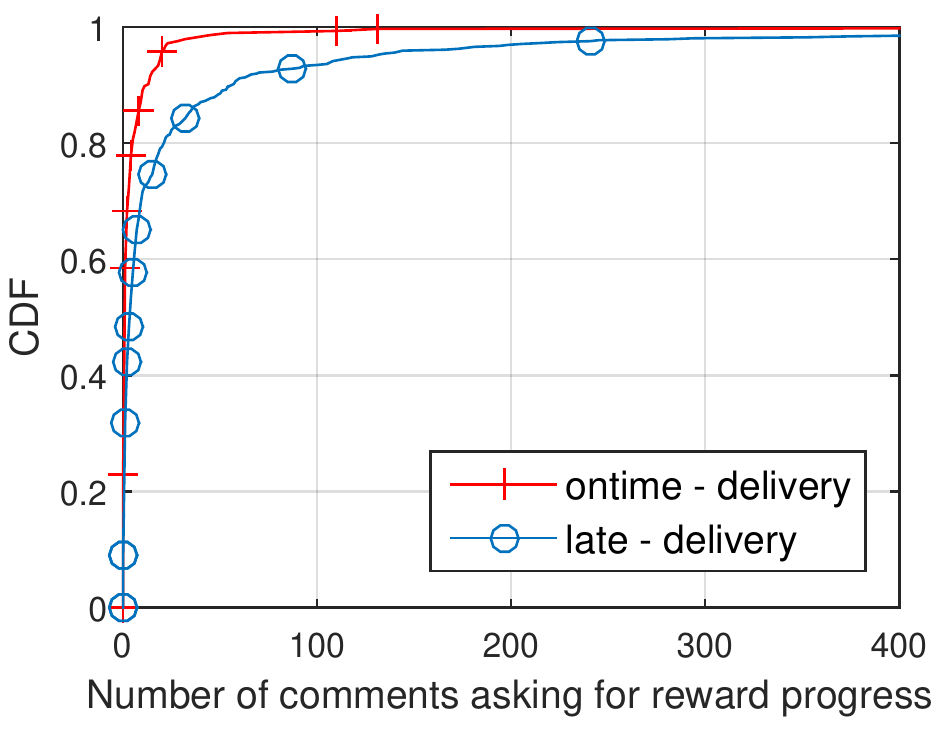}
	}
	\caption{Activeness of creators and backers in on-time and late delivery projects during the delivery duration. They in two groups behaved differently. *,**,***, indicate p-value $< 0.05$, $< 0.01$, and $< 0.001$, respectively.}
	\label{fig:HowOften-HowFast-Creator-Update}
\end{figure*}

\smallskip
\noindent\textbf{Activeness of creators and backers during delivery duration:} First, we measured the activeness of creators in two aspects: (1) the average time interval between two consecutive updates during the delivery duration; and (2) the responsiveness of creators regarding a backer's comment.

Figure~\ref{fig:HowFast-Creator-Update} shows the average update time interval of on-time and late delivery projects. Creators of the on-time delivery projects posted updates with shorter time interval than creators of the late delivery projects. Creators of on-time and late delivery projects had significantly different behavior with p-value $<$ 0.001.

Backers often posted questions, as comments, regarding the status of rewards during the delivery duration. Our hypothesis is that creators of the on-time delivery projects would respond more quickly than creators of late delivery projects, showing they were paying more attention to the backers. Figure~\ref{fig:HowFast-Creator-Comment} shows that our hypothesis is correct. Creators of the on-time delivery projects had significantly different behavior compared with creators of the late delivery project (p-value $<$ 0.05).

Next, we measured the activeness of backers during the delivery duration. Did backers in on-time delivery projects behave differently from ones in late delivery projects in terms of posting more or less number of questions? Figure~\ref{fig:CDF-BackerAsk4RewardProgress} shows the number of questions posted by backers in on-time and late delivery projects. It makes sense that backers in late delivery projects posted more questions than backers in on-time delivery projects (p-value $<$ 0.001).

\smallskip
\noindent\textbf{Linguistic patterns:} Did creators in on-time and late delivery projects have different linguistic usage? To answer this question, we analyzed creators' updates posted during the delivery duration in terms of linguistic patterns. In particular, we used Linguistic Inquiry and Word Count (LIWC) dictionary, a standard approach for mapping text to psychologically meaningful categories in various research areas \cite{lee2014will}. LIWC-2007 defines 64 categories for each of several languages (e.g., English, German, Dutch). Each category contains several dozens to hundred of words or stems.

Given each project's updates which were posted during \emph{the reward delivery phase}, we measured the linguistic usage of the creator in 64 LIWC categories by computing his score with regard to each category in LIWC dictionary. First, we grouped all the updates of each project into one big document. Next, we removed all stop words and counted the total number of words \textit{N} in the document. Then, we counted the number of words in the document overlapped with the words or matched the stems in each category \textit{i} in LIWC dictionary, denoted by \textit{$C_i$}. Finally, we computed the creator's score of a LIWC category \textit{i} as \textit{$C_i/N$}.

We performed the two-sample t-test to discover the LIWC categories in which two distributions had a significant difference in the mean of score. Since we performed simultaneous tests against 64 LIWC categories, we applied Bonferroni correction and assigned $\alpha$ as 0.00078 (=0.05/64). Finally, we observed 12 categories that had the significant difference of linguistic usage of creators in on-time and late delivery projects. Top 3 distinguishing linguistic categories were \emph{positive emotion}, \emph{affective process} and \emph{cognitive processes}. Interestingly, creators in on-time delivery projects used words related to \emph{positive emotion} (e.g., love, sweet) and \emph{affective processes (e.g., happy)} more than creators in late delivery projects. However, creators in on-time delivery projects used words related to \emph{Cognitive processes} (e.g., cause, know, ought) less than creators in late delivery projects. This makes sense since creators in late delivery projects usually posted updates about reasons why their reward production or delivery process was delayed.

We also analyzed backers' comments to understand their linguistic usage via the same process and found 5 categories with significant difference: \emph{positive emotion}, \emph{negative emotion}, \emph{anxiety}, \emph{cognitive processes} and \emph{insight}. In particular, backers in on-time delivery projects used words related to \emph{positive emotion} more, and related to \emph{negative emotion} and \emph{cognitive processes} less than backers in late delivery projects. This result is consistent with creators' linguistic patterns.

\section{Conclusion}
It is crucial for creators to deliver rewards to backers on time as it keeps and fosters the trust between them. As the first step to achieve trustworthy crowdfunding platforms, we analyzed characteristics of on-time and late reward delivery projects. Particularly, we analyzed project-based characteristics, creator-based characteristics, activeness-based and linguistic-based characteristics of creators and backers during \emph{the reward delivery phase}. We found that less complicated projects with more active response to backers had a higher chance to ship rewards on time.


\section{Acknowledgment}
This work was supported in part by NSF grant CNS-1553035. Any opinions, findings and conclusions or recommendations expressed in this material are the author(s) and do not necessarily reflect those of the sponsors.

{
\small
\bibliographystyle{aaai}
\bibliography{sigproc}  
}

\end{document}